\documentstyle[preprint,aps]{revtex}

\begin{document}

\draft

\title{
Sea contributions to spin 1/2 baryon structure,\\
magnetic moments, and spin distribution
}

\author{
V.~Gupta\cite{email1},\
R.~Huerta\cite{email2},\
and
G.~S\'anchez-Col\'on\cite{email3}.
}
\address{
Departamento de F\'{\i}sica Aplicada. \\
Centro de Investigaci\'on y de Estudios Avanzados del IPN.\\
Unidad M\'erida. \\
A.P. 73, Cordemex 97310, M\'erida, Yucat\'an, MEXICO. \\
}

\date{March 7, 1996}

\maketitle

\begin{abstract}
We treat the baryon as a composite system made out of a \lq\lq core" of three
quarks (as in the standard quark model) surrounded by a \lq\lq sea" (of gluons
and $q\bar{q}$-pairs) which is specified by its total quantum numbers like
flavor, spin and color.
Specifically, we assume the sea to be a flavor octet with spin 0 or 1 but no
color.
The general wavefunction for spin 1/2 baryons with such a sea component is
given.
Application to the magnetic moments is considered.
Numerical analysis shows that a scalar (spin 0) sea with an
admixture of a vector (spin 1) sea can provide very good fits to the magnetic
moment data {\em using experimental errors}.
Our best fit automatically gives $g_A/g_V$ for neutron beta decay in agreement
with data.
This fit also gives reasonable values for the spin distributions of the proton
and neutron.
\end{abstract}


\section{Introduction}
\label{i}

Attempts to understand the static properties of hadrons in the framework of
the standard quark model (SQM) have had limited success.
The belief that valence quarks were responsible for the spin of the proton has
been shattered by recent experiments\cite{1}.

The naive valence quark picture of hadron structure is a simplification which
does not properly take into account the fact that quarks interact through
color forces mediated by vector gluons.
The existence of the quark-gluon interaction, in QCD, implies that a hadron
should be viewed as consisting of valence quarks surrounded by a \lq\lq sea"
which contains gluons and virtual quark-antiquark ($q\bar{q}$) pairs.
Deep inelastic lepton-nucleon scattering has shown the existence of a sea
component and its importance for nucleon structure functions.
It is thus necessary to understand how this sea contributes to the baryon spin
and other low energy properties.

Several authors\cite{2,3,4,5,6,7} have studied the effect of the sea
contributions on the hadron structure and  the static properties of baryons.
Some consider the sea as a single gluon or a $q\bar{q}$-pair.
However the sea, in general, consists of any number of gluons and
$q\bar{q}$-pairs.
In this paper we \lq\lq model" the general sea by its total quantum numbers
(flavor, spin and color) which are such that the sea wave function when
combined with the valence quark wave function gives the desired quantum
numbers for the physical hadron.
In particular, we explore the consequences of a \lq\lq sea" with flavor and
spin but no color\cite{6,7} for the low energy properties of the spin 1/2
baryon
octet $(p,n,\Lambda,\dots)$.

For simplicity, we consider a flavor octet sea with spin 0 and spin 1.
The physical baryon wavefunction incorporating the sea is used to calculate the
baryon magnetic moments, understanding of which is our primary motivation.
We find that a scalar (spin 0) sea and a vector
(spin 1) sea described by two and one parameters respectively gives
a very good fits to the magnetic moment data.
These fits give us numerical predictions for the spin distributions of the
nucleons and $g_A/g_V$ for neutron beta decay.

In Sec.~\ref{ii}, we discuss the wavefunctions for the physical baryons
constructed from the valence quarks and our model for the sea.
In Sec.~\ref{iii}, we obtain the magnetic moments from the modified
wavefunction and a general discussion of the results for them is given in
Sec.~\ref{results}.
Sec.~\ref{iv} gives the consequences of the fits for the spin distributions.
Our fits automatically predict $g_A/g_V$; the axial vector weak decay constant
for neutron beta decay.
Use of this datum as a constraint on the fits is briefly discussed in
Sec.~\ref{vp}.
Sec.~\ref{v} gives a summary and discussion.

\section{Spin 1/2 octet baryon wavefunctions with sea}
\label{ii}

For the lowest-lying baryons, in the SQM, the three valence quarks are taken
to be in a relative $S$-wave states.
The flavor-spin wavefunction of the three quarks is totally symmetric while
the color wavefunction is totally antisymmetric to give a color singlet
baryon.
For the $SU(3)$ flavor octet spin 1/2 baryons we denote this SQM or $q^3$
wavefunction by $\tilde{B}({\bf 8},1/2)$, the argument 1/2 refers to spin.
These $J^P=\frac{1}{2}^+$ states are denoted by $\tilde{p}$,
$\tilde{\Sigma}^{+}$, etc.

Representing the baryons by this wavefunction is at best a first-order
approximation.
In reality, since quarks interact, the baryons contain a \lq\lq sea" of gluons
and virtual $q\bar{q}$-pairs in addition to the valence quarks.
The important question is how to take into account this general sea.
We take this general sea to be described by wavefunctions which are specified
by the total flavor, spin, and color quantum numbers of the sea.
We picture the physical baryon as a $q^3$ \lq\lq core" (described by SQM
wavefunction $\tilde{B}({\bf 8},1/2)$) sorrounded by a sea.

The physical baryon octet states, denoted $B({\bf 8},1/2)$ are obtained by
combining the \lq\lq core" wavefunction $\tilde{B}({\bf 8},1/2)$ with the sea
wavefunction with specific properties given below.

We assume the sea is a color singlet but has flavor and spin properties which
when combined with those of the core baryons $\tilde{B}$ give the desired
properties of the physical baryon $B$.
We further assume that there is no relative orbital angular momentum between
the core and the sea.
Since both the physical and core baryon have $J^P=\frac{1}{2}^+$, this implies
that the sea has even parity and spin 0 or 1.
The spin 0 and 1 wavefunction for the sea are denoted by $H_0$ and $H_1$,
respectively.
We also refer to a spin 0 (1) sea as a scalar (vector) sea.
For $SU(3)$ flavor we assume the sea has a $SU(3)$ singlet component and an
octet component described by wavefunctions $S({\bf 1})$ and $S({\bf 8})$,
respectively.
The color singlet sea in our model is thus described by the wavefunctions
$S({\bf 1})H_0$, $S({\bf 1})H_1$, $S({\bf 8})H_0$, and $S({\bf 8})H_1$.

The $SU(3)$ symmetric and spinless sea component implicit in SQM is described
by $S({\bf 1})H_0$.
Such a color singlet sea would require at least two gluons or a
$q\bar{q}$-pair.
The sea described by the wavefunctions $S({\bf 1})H_1$, $S({\bf 8})H_0$,
and $S({\bf 8})H_1$ require a minimum of one $q\bar{q}$-pair.
The flavor and spin quantum numbers of the sea considered here are simple or
minimal in the sense that they require only one $q\bar{q}$-pair.
However, the color singlet sea described by the above wavefunctions can, in
general, have any number of $q\bar{q}$-pairs and gluons consistent with its
total flavor and spin quantum numbers.

We thus represent the physical baryon states as a superposition of different
combinations of the core baryons with the sea wavefunctions specified above,
namely

\begin{eqnarray}
B(1/2)
&\sim&
\tilde{B}({\bf 8},1/2)H_{0}S({\bf 1})+
\tilde{B}({\bf 8},1/2)H_{0}S({\bf 8})
\nonumber\\
&&
+\tilde{B}({\bf 8},1/2)H_{1}S({\bf 1})+
\tilde{B}({\bf 8},1/2)H_{1}S({\bf 8}).
\label{uno}
\end{eqnarray}

\noindent
The flavor and spin quantum numbers of the core baryons $\tilde{B}$ and the
sea (in each term) combine to give the quantum numbers of the physical baryon
$B$.
The baryon $B$ is thus part of the time just $\tilde{B}$ with an inert sea
(first term in Eq.~(\ref{uno})) and part of the time $\tilde{B}$ plus sea with
flavor and spin (last three terms in Eq.~(\ref{uno})).

Because the sea has flavor, Eq.~(\ref{uno}) implies that a baryon $B(Y,I)$,
with given isospin $I$ and hypercharge $Y$, will contain core baryons
$\tilde{B}$ with different $I$ and $Y$ thus the $B(Y,I)$ will have a very
different quark content from the corresponding SQM-state $\tilde{B}(Y,I)$.
For example, the physical proton $p$ will contain terms involving $\tilde{p}$,
$\tilde{\Lambda}$, $\tilde{\Sigma}^+$, etc. plus sea and have non-zero strange
quark content unlike $\tilde{p}$ in SQM (see Eq.~(\ref{siete})).
We now go on to formulate the baryon wavefunction more precisely.

The total flavor-spin wavefunction of a spin up ($\uparrow$) physical baryon
which consists of 3 valence quarks and a sea component (as discussed above)
can be written schematically as

\begin{eqnarray}
B(1/2\uparrow)
&=&
\tilde{B}({\bf 8},1/2\uparrow)H_{0}S({\bf 1})+
b_{{\bf 0}}\left[\tilde{B}({\bf 8},1/2)
\otimes H_{1}\right]^{\uparrow}S({\bf 1})
\nonumber\\
&&
+\sum_{N}a(N)\left[\tilde{B}({\bf 8},1/2\uparrow)H_{0}
\otimes S({\bf 8})\right]_{N}
\label{dos}\\
&&
+\sum_{N}b(N)\left\{[\tilde{B}({\bf 8},1/2)\otimes H_{1}]^{\uparrow}
\otimes S({\bf 8})\right\}_{N}.
\nonumber
\end{eqnarray}

\noindent
The normalization not indicated here is discussed later.
The first term is the usual $q^{3}$-wavefunction of the SQM (with a trivial
sea) and the second term (coefficent $b_0$) comes from spin-1 (vector) sea
which combines with the spin 1/2 core baryon $\tilde{B}$ to a
spin 1/2$\uparrow$ state.
So that,

\begin{equation}
\left[\tilde{B}({\bf 8},1/2)\otimes H_{1}\right]^{\uparrow} =
\sqrt{\frac{2}{3}}\tilde{B}({\bf 8},1/2\downarrow)H_{1,1} -
\sqrt{\frac{1}{3}}\tilde{B}({\bf 8},1/2\uparrow)H_{1,0}.
\label{tres}
\end{equation}

\noindent
In both these terms the sea is a flavor singlet.
The third (fourth) term in Eq.~(\ref{dos}) contains a scalar (vector) sea
which transforms as a flavor octet.
The various $SU(3)$ flavor representations obtained from
$\tilde{B}({\bf 8})\otimes S({\bf 8})$ are labelled by
$N={\bf 1,8_{F},8_{D},10,\bar{10},27}$.
As it stands, Eq.~(\ref{dos}) represents a spin 1/2$\uparrow$ baryon which is
not {\em a pure flavor octet} but has an admixture of other $SU(3)$
representations weighted by the unspecified constants $a(N)$ and $b(N)$.
It will be a flavor octet if $a(N)=b(N)=0$ for $N={\bf 1,10,\bar{10},27}$.
The color wavefunctions have not been indicated as the three valence quarks in
the core $\tilde{B}$ and the sea (by assumption) are in a color singlet state.

For our applications we adopt the phenomenological wavefunction given in
Eq.~(\ref{dos}), where the physical spin 1/2 baryons have admixtures of flavor
$SU(3)$ determined by the coefficients $a(N)$ and $b(N)$,
$N={\bf 1,10,\bar{10},27}$.
As we shall see, such a wavefunction which respects the isospin and
hypercharge properties of the usual spin 1/2 baryon states is general enough
to provide an excellent fit to the magnetic moments data.
Surprisingingly, only 3 or 4 of the thirteen parameters in Eq.~(\ref{dos}) are
needed for this purpose.
For the moment we discuss the general wavefunction in Eq.~(\ref{dos}) as it is.
Incidentally, such a wavefunction could arise in general since we know flavor
is broken by mass terms in the QCD Lagrangian.

The sea isoespin multiplets contained in the octet $S({\bf 8})$ are denoted as

\begin{equation}
(S_{\pi^+},S_{\pi^0},S_{\pi^-}),\qquad
(S_{K^+},S_{K^0}),\qquad
(S_{\bar{K}^0},S_{K^-}),\quad
\hbox{and}\quad S_{\eta}.
\label{cuatro}
\end{equation}

\noindent
The suffix on the components label the isospin and hypercharge quantum numbers.
Note, the familiar pseudoscalar mesons are used here as subscripts only to
label the flavor quantum numbers of the sea states.
All the components of $S({\bf 8})$ have $J^P=0^+$ as mentioned earlier.
For example, $S_{\pi^+}$ has $I=1,I_3=1$ and $Y=0$; $S_{K^-}$ has
$I=1/2,I_3=-1/2$, and $Y=-1$; etc.
These flavor quantum numbers when combined with those of the three valence
quarks states $\tilde{B}$ will give the observed $I$, $I_3$, and $Y$ for the
physical states $B$.
The flavor combinations in the third and fourth terms in Eq.~(\ref{dos}) imply
that the
physical states $B(Y,I,I_3)$ are expressed as a sum of products of
$\tilde{B}(Y,I,I_3)$ and the sea components $S(Y,I,I_3)$, weighted by some
coefficients $\alpha_i$ which are linear combinations of the coefficients
$a(N)$ and $b(N)$.
Schematically, the flavor content of the third and fourth terms in
Eq.~(\ref{dos}) is of the form (suppressing $I_3$)

\begin{equation}
B(Y,I)=
\sum_i \alpha_i(Y_1,Y_2,I_1,I_2)\left[\tilde{B}(Y_1,I_1)S(Y_2,I_2)\right]_i
\label{cinco}
\end{equation}

\noindent
where the sum is over all $Y_i$, $I_i$, ($i=1,2$); such that: $Y=Y_1+Y_2$ and
$\bf I=I_1+I_2$.
The flavor content of $B(Y,I,I_3)$ in terms of $\tilde{B}(Y,I,I_3)$ and sea
components are given in Table~\ref{tabla1}.
The corresponding coefficients $\bar{\beta}_i$, $\beta_i$, etc. expressed in
terms of the coefficients $a(N)$ (for the scalar sea) are recorded in
Table~\ref{tabla2}.
In Table~\ref{tabla1} we have denoted $\tilde{B}(Y,I,I_3)$ and $S(Y,I,I_3)$ by
appropiate symbols, e.g., $\tilde{B}(1,1,1/2)$ by $\tilde{p}$, $S(0,1,1)$ by
$S_{\pi^+}$, etc.
Since the flavor content of the fourth term with vector sea is the same as for
the scalar sea, the contribution of the fourth term in Eq.~(\ref{dos}) to the
physical baryon state can be obtained by using Eq.~(\ref{tres}) and
Tables~\ref{tabla1} and \ref{tabla2} with the replacement
$a(N)\rightarrow b(N)$ for $N={\bf 1,8_F,8_D,10,\bar{10},27}$.
For later use, the coefficients obtained by changing $a(N)\rightarrow b(N)$ in
$\bar{\beta}_i$, $\beta_i$, $\gamma_i$, and $\delta_i$ will be denoted by
$\bar{\beta'}_i$, $\beta'_i$, $\gamma'_i$, and $\delta'_i$.
In Tables~\ref{tabla1} and \ref{tabla2} for the reduction of
$\tilde{B}({\bf 8})\otimes S({\bf 8})$ into various $SU(3)$ representations we
have followed the convention used by Carruthers~\cite{carruthers}.

The normalization of the physical baryons wavefunction in Eq.~(\ref{dos}) can
be obtained by using
$\langle H_i | H_j\rangle  = \delta_{ij}$,
$\langle \tilde{B}(Y,I,I_3)|\tilde{B}(Y',I',I'_3)\rangle =
\langle S(Y,I,I_3)|S(Y',I',I'_3)\rangle =
\delta_{YY'}\delta_{II'}\delta_{I_3I'_3}$.
However, it should be noted that the normalization are different, in general,
for each $B(Y,I)$ state.
This is because not all $a(N)$ and $b(N)$ contribute to a given
$(Y,I)$-multiplet as is clear from Tables~\ref{tabla1} and \ref{tabla2}.
For example, $a({\bf 1})$ and $b({\bf 1})$ contribute only to $\Lambda$ while
$a({\bf 10})$ and $b({\bf 10})$ do not contribute to the nucleon states.
Denoting by $N_1$, $N_2$, $N_3$, and $N_4$, the normalization constants for
the $(p,n)$, $(\Xi^0,\Xi^-)$, $(\Sigma^{\pm},\Sigma^0)$, and $\Lambda$
isospin multiplets, one has

\begin{mathletters}
\label{seis}
\begin{equation}
N^2_1 = N^2_0 + a^2({\bf\bar{10}}) + b^2({\bf\bar{10}}),
\end{equation}
\begin{equation}
N^2_2 = N^2_0 + a^2({\bf 10}) + b^2({\bf10}),
\end{equation}
\begin{equation}
N^2_3 = N^2_0 + \sum_{N={\bf 10,\bar{10}}}[a^2(N) + b^2(N)],
\end{equation}
\begin{equation}
N^2_4 = N^2_0 + a^2({\bf 1}) + b^2({\bf 1}),
\end{equation}

\noindent
where,

\begin{equation}
N^2_0 = 1 + b^2_0 + \sum_{N={\bf 8_D,8_F,27}}[a^2(N) + b^2(N)].
\end{equation}
\end{mathletters}

\noindent
For example, using Tables~\ref{tabla1} and \ref{tabla2}, and Eqs.~(\ref{seis}),
the physical spin-up proton state as given by Eq.~(\ref{dos}) is

\begin{eqnarray}
N_1 |p\uparrow\rangle
&=&
|\tilde{p}\uparrow\rangle H_0S({\bf 1})+b_0|
(\tilde{p}\otimes H_1)^{\uparrow}\rangle S({\bf 1})
\nonumber\\
&&
+ \bar{\beta}_1 |\tilde{p}\uparrow\rangle  S_{\eta}
+ \bar{\beta}_2 |\tilde{\Lambda}\uparrow\rangle  S_{K^+}
+ \bar{\beta}_3 |(\tilde{N}\uparrow S_{\pi})_{1/2,1/2}\rangle
+ \bar{\beta}_4 |(\tilde{\Sigma}\uparrow S_K)_{1/2,1/2}\rangle
\label{siete}\\
&&
+ \bar{\beta'}_1 |(\tilde{p}\otimes H_1)^{\uparrow}\rangle  S_{\eta}
+ \bar{\beta'}_2 |(\tilde{\Lambda}\otimes H_1)^{\uparrow}\rangle  S_{K^+}
\nonumber\\
&&
+ \bar{\beta'}_3 |((\tilde{N}\otimes H_1)^{\uparrow} S_{\pi})_{1/2,1/2}\rangle
+ \bar{\beta'}_4 |((\tilde{\Sigma}\otimes H_1)^{\uparrow} S_K)_{1/2,1/2}\rangle
\nonumber,
\end{eqnarray}

\noindent
where $(\tilde{B}\otimes H_1)^{\uparrow}$ are given in Eq.~(\ref{tres}) and
$\bar{\beta'}_1=
(3b({\bf 27})-b({\bf 8_D})+(b({\bf 8_F})+b({\bf \bar{10}}))/2)/\sqrt{20}$,
and so on.
Other baryon wavefunctions will have a similar structure.
Also, $(\tilde{N}\uparrow S_{\pi})_{1/2,1/2}$
($(\tilde{\Sigma}\uparrow S_{K})_{1/2,1/2}$) stand for the $I=I_3=1/2$
combination of the $I=1/2$ $\tilde{N}$ ($S_K$) and $I=1$ $S_{\pi}$
($\tilde{\Sigma}$) multiplets.

For any operator $\hat{O}$ which depends only on quarks, the matrix elements
are easily obtained using the ortogonality of the sea components.
Clearly $\langle p\uparrow|\hat{O}|p\uparrow\rangle $ will be a linear
combination of the matrix elements
$\langle \tilde{B}\uparrow|\hat{O}|\tilde{B'}\uparrow\rangle$ (known from
SQM) with coefficients which depend on the coefficients in the wavefunction.

For applications, we need the quantities $(\Delta q)^B$, $q=u,d,s$; for each
spin-up baryon $B$.
These are defined as

\begin{equation}
(\Delta q)^B =
n^B(q\uparrow)-n^B(q\downarrow)+n^B(\bar{q}\uparrow)-n^B(\bar{q}\downarrow),
\label{ocho}
\end{equation}

\noindent
where $n^B(q\uparrow)$ ($n^B(q\downarrow)$) are the number of spin-up
(spin-down) quarks of flavor $q$ in the spin-up baryon $B$.
Also, $n^B(\bar{q}\uparrow)$ and $n^B(\bar{q}\downarrow)$ have a similar
meaning for antiquarks.
However, these are zero as there are no explicit antiquarks in the
wavefunctions given by Eq.~(\ref{dos}).
The expressions for $(\Delta q)^B$ are given in Table~\ref{tabla3} in terms of
the coefficients $b_0$, $\bar{\beta}_i$, $\bar{\beta'}_i$, etc.
Note that the terms involving $b_0$, $\bar{\beta'}_i$, $\beta'_i$,
$\gamma'_i$, and $\delta'_i$ are multiplied by the factor $-1/3$ which comes
from Eq.~(\ref{tres}) on taking the matrix element of the operator
$\hat{\Delta q}$.
The expressions for $(\Delta q)^B$ reduce to the SQM values if there is no sea
contribution, that is, $b_0=0$, $a(N)=b(N)=0$,
$N={\bf 1,8_F,8_D,10,\bar{10},27}$.
Moreover, the total spin $S_Z$ of a baryon is given by
$S^B_Z = (1/2) \sum_q (\Delta q)^B + (\Delta(\hbox{sea}))^B$,
where the second term represents the spin carried by the sea and depends
solely on $b_0$ and $b(N)$'s, the coefficients determining the vector sea.
For $S^B_Z=1/2$, we expect $\sum_q (\Delta q)^B = 1$ for a purely scalar sea,
i.e., when $b_0$ and all $b(N)$'s are zero.
This is indeed true for each baryon as can be seen from Table~\ref{tabla3}.
There are three $(\Delta q)^B$ $(q=u,d,s)$ for each $(Y,I)$-multiplet.
These twelve quantities and $(\Delta q)^{\Sigma^0\Lambda}$ are given in terms
of the thirteen parameters of Eq.~(\ref{dos}) as our spin 1/2 baryons do not
belong to a definite representation of $SU(3)$.
To obtain a flavor octet physical baryon one restricts $N$ to ${\bf 8_F}$ and
${\bf 8_D}$ in Eq.~(\ref{dos}), that is, put $a(N)=b(N)=0$ for
$N={\bf 27,10,\bar{10},1}$, so that the twelve $(\Delta q)^B$ are given in
terms of five parameters $b_0$, $a(N)$, $b(N)$ with $N={\bf 8_F,8_D}$.
It is clear, in this case, that our wavefunction provides a model for spin 1/2
baryons which is more general than the phenomenological model considered by
some authors\cite{9} recently to fit the baryon magnetic moments.
These authors take the three quantities $(\Delta q)^p$ $(q=u,d,s)$ as
parameters to be determined from data but use flavor $SU(3)$ to express all
the other $(\Delta q)^B$ in terms of the $(\Delta q)^p$.
In our case the various $(\Delta q)^B$ are not simply related by flavor $SU(3)$
because of the non-trivial flavor properties of the sea and thus provides an
explicit and very different model for the baryons.

\section{Application to magnetic moments}
\label{iii}

We assume the baryon magnetic moment operator $\hat{\mu}$ to be expressed
solely in terms of quarks as is usual in the quark model.
So that $\hat{\mu}=\sum_q (e_q/2m_q) \sigma^q_Z$ ($q=u,d,s$).
It is clear from Eq.~(\ref{dos}) that $\mu_B=\langle B|\hat{\mu}|B\rangle $
will be a linear combination of $\mu_{\tilde{B}}$ and
$\mu_{\tilde{\Sigma}^0\tilde{\Lambda}}$ weighted by the coefficients which
depend on $b_0$, $a(N)$'s, and $b(N)$'s.
The magnetic moments $\mu_{\tilde{B}}$ and the transition moment
$\mu_{\tilde{\Sigma}^0\tilde{\Lambda}}$ (for the core baryons) are given in
terms of the quark magnetic moments $\mu_q$ as per SQM.
For example,
$\mu_{\tilde{p}}=(4\mu_u-\mu_d)/3$,
$\mu_{\tilde{\Lambda}}=\mu_s$,
$\mu_{\tilde{\Sigma}^0\tilde{\Lambda}}=(\mu_u-\mu_d)/\sqrt{3}$,
etc.
Consequently, all the magnetic moments and the $\Sigma^0\rightarrow\Lambda$
transition magnetic moment in our model can be written simply as

\begin{mathletters}
\label{nueve}
\begin{eqnarray}
\mu_B &=& \sum_q (\Delta q)^B \mu_q,\ \ \ \ (q=u,d,s);
\\
\mu_{\Sigma^0\Lambda} &=&
\sum_q (\Delta q)^{\Sigma^0\Lambda} \mu_q,\ \ \ \ (q=u,d);
\end{eqnarray}
\end{mathletters}

\noindent
where the $(\Delta q)^B$ and $(\Delta q)^{\Sigma^0\Lambda}$ are given in
Table~\ref{tabla3} and $B=p,n,\Lambda,\dots$.

A class of models\cite{9} have been recently considered in which the magnetic
moments were expressed in terms of $\mu_q$ and $(\Delta q)^p$ ($q=u,d,s$)
without giving an explicit wavefunction.
Interestingly, Eqs.~(\ref{nueve}) have the same general structure except that
here the twelve $(\Delta q)^B$ and $(\Delta q)^{\Sigma^0\Lambda}$ are not
related but depend on thirteen parameters, namely, $b_0$ and six $b(N)$'s for
the vector sea and the six $a(N)$'s for the scalar sea.
Despite, our general wavefunction, the isospin sum rule\cite{11}

\begin{equation}
\mu_{\Sigma^0}=\frac{1}{2}(\mu_{\Sigma^+}+\mu_{\Sigma^-})
\label{diez}
\end{equation}

\noindent
holds.
This is because $\hat{\mu}$ transforms as $(I=0) \oplus (I=1)$ in isospin
space and the wavefunction of Eq.~(\ref{dos}) respects isospin thus giving
$(\Delta q)^{\Sigma^0}=((\Delta q)^{\Sigma^+}+(\Delta q)^{\Sigma^-})/2$
(see Table~\ref{tabla3}).

\paragraph*{\bf Discussion of the case when physical baryons form a flavor
octet.}

If we require the physical baryon states given by Eq.~(\ref{dos}) to
transform as an $SU(3)$ octet (i.e., put $a(N)=b(N)=0$,
$N={\bf 1,10,\bar{10},27}$) then the seven $\mu_B$'s and
$\mu_{\Sigma^0\Lambda}$ depend non-linearly on five parameters
($b_0$, $b({\bf 8_F})$, $b({\bf 8_D})$, $a({\bf 8_F})$, and $a({\bf 8_D})$)
of the wavefunction and the three $\mu_q$'s.
Even so, three sum rules, namely

\begin{mathletters}
\label{once}
\begin{eqnarray}
\mu_p - \mu_n &=&
(\mu_{\Sigma^+} - \mu_{\Sigma^-}) - (\mu_{\Xi^0} - \mu_{\Xi^-})
\\
((4.70589019\pm 5\times 10^{-7})\ \mu_N)
&&
((4.217\pm 0.031)\ \mu_N)
\end{eqnarray}
\end{mathletters}

\begin{mathletters}
\label{doce}
\begin{eqnarray}
-6\mu_{\Lambda} &=&
\mu_{\Sigma^+} + \mu_{\Sigma^-} - 2(\mu_p + \mu_n + \mu_{\Xi^0} + \mu_{\Xi^-})
\\
((3.678\pm 0.024)\ \mu_N)
&&
((3.340\pm 0.039)\ \mu_N)
\end{eqnarray}
\end{mathletters}

\begin{mathletters}
\label{docep}
\begin{eqnarray}
2\sqrt{3}\mu_{\Sigma^0\Lambda} &=&
2(\mu_p - \mu_n) - (\mu_{\Sigma^+} - \mu_{\Sigma^-})
\\
((5.577\pm 0.277)\ \mu_N)
&&
((5.794\pm 0.027)\ \mu_N)
\end{eqnarray}
\end{mathletters}

\noindent
emerge.
These have been noted earlier in the context of other models\cite{6,9}.
The values of the two sides taken from data~\cite{10} are shown in parentheses.
The reason these sum rules hold despite the number of parameters is because
they are a consequence of flavor $SU(3)$ since baryons form a $SU(3)$ octet
and $\hat{\mu}$ transforms as ${\bf 1}\oplus{\bf 8}$.
However, as can be seen, the first two $SU(3)$ sum rules are not well satisfied
experimentally.
To avoid them one could modify the $SU(3)$ transformation properties of
$\hat{\mu}$ or the baryons.
A group-theoretic analysis with the most general $\hat{\mu}$ which would
contribute to the magnetic moments of an octet was done by Dothan\cite{12}
over a decade ago.
Such a $\hat{\mu}$ could arise from $SU(3)$ breaking effects.
However, several authors\cite{13} have considered models in which they modify
the baryon wavefunction.
In our approach, we keep $\hat{\mu}$ as given by the quark model but modify
the baryon wavefunction by taking a sea with flavor and spin into account as
in Sec.~\ref{ii}.

SQM has three parameters $\mu_q$ ($q=u,\ d,\ s$) the quark magnetic moments in
nuclear magnetons $\mu_N$.
A fit using experimental errors gives $\chi^2/\mbox{DOF}=1818/5$ with
$\mu_u=1.8517$, $\mu_d=-0.9719$, $\mu_s=-0.7013$.
These values for $\mu_q$ differ from the values given in Ref.\cite{10}, since
there $\mu_p$, $\mu_n$, and $\mu_{\Lambda}$ are used as inputs.

The situation improves a little for a pure octet physical baryon with scalar
and vector sea described by $a({\bf 8_F})$, $a({\bf 8_D})$, $b_0$,
$b({\bf 8_F})$, and $b({\bf 8_D})$.
These five sea parameters enter Eqs.~(\ref{nueve}) only through the three
combinations given by $(\Delta q)^p$.
Hence, the 3 sum rules in Eqs.~(\ref{once})-(\ref{docep}).
For {\em experimental errors }with $\mu_q$ also as parameters one
obtains~\footnote{Due to the form of Eqs.~(\ref{nueve}) there are only five
effective parameters~\cite{9}.}
$\chi^2/\mbox{DOF}=652/3$.
Most of the contribution to $\chi^2$ comes from a poor fit to $\mu_{\Sigma^+}$,
$\mu_{\Sigma^-}$, and $\mu_{\Xi^0}$.
This is a clear indication that admixture of other $SU(3)$ representations in
our wavefunction need to be considered.

As noted above, even with both scalar and vector sea present, the poorly
satisfied $SU(3)$ sum rules Eqs.~(\ref{once})-(\ref{docep}) will hold as
long the physical baryon is restricted to be an octet.
This means we must include $SU(3)$ breaking effects in the baryon wavefunction
by considering non-zero $a(N)$ and or $b(N)$ with $N={\bf 1,10,\bar{10},27}$.

\section{Results}
\label{results}

In making our fits we have used {\em experimental errors }as given by Particle
Data Group\cite{10}.
This is in contrast to many authors who use \lq\lq theoretical errors" of the
order of a few percent or more to fit the data.
In actual fact the experimental errors are much smaller.
Furthermore, we keep in mind that the constituent quark masses are
$m_u,\ m_d\approx 300{\rm MeV}$, and $m_s\approx 500{\rm Mev}$, so we expect
$\mu_u\cong -2\mu_d> 0$ and $\mu_s\cong 0.6 \mu_d$.
Current quark masses would give a very different numerical range for the
ratios $\mu_u/\mu_d$ and $\mu_s/\mu_d$.
Also, if the core baryon contribution is dominant then the parameters
determining the sea should be small compared to unity.
Furthermore, for a dominantly $SU(3)$ octet physical baryon $b_0$,
$a({\bf 8_F})$, $a({\bf 8_D})$, $b({\bf 8_F})$, and $b({\bf 8_D})$ should be
larger than the other parameters in the wave function.

To get a feeling for how the sea contributes we did extensive and systematic
numerical analysis separately for the three cases: pure scalar (spin 0)
sea, pure vector (spin 1) sea, and scalar plus vector sea.
In all the fits, in addition to the sea parameters, $\mu_q$ were treated as
parameters.

\paragraph*{\bf Scalar sea.}
\label{iiia}

In general, here there are six parameters $a(N)$'s,
$N={\bf 1, 8_F, 8_D, 10, \bar{10}, 27}$, in the wavefunction and the
three $\mu_q$'s.
These nine parameters provide a perfect fit with $\chi^2=1.5\times 10^{-4}$.
This clearly means that the scalar sea contribution modifies the values of
$(\Delta q)^B$ in the right direction for a fit to the baryon magnetic moments.
However, a seven parameter fit with
$a({\bf 1})=0.0625$,
$a({\bf 8_D})=-0.1558$,
$a({\bf 8_F})=0.1896$,
$a({\bf 10})=0.4297$,
and
$\mu_u=1.8589$,
$\mu_d=-0.9988$,
and
$\mu_s=-0.6530$,
provides an excellent fit with $\chi^2/\mbox{DOF}=0.838$.
The $\mu_q$ (in units of $\mu_N$) imply for the quark masses the values
$m_u=336.49\ {\rm MeV}$,
$m_d=313.12\ {\rm MeV}$,
and
$m_s=478.94\ {\rm MeV}$,
which are in accord with the constituent quark model.
A noteworthy six parameter fit with $\chi^2/\mbox{DOF}=5.60/2$ is given by
$a({\bf 8_D})=-0.2262$,
$a({\bf 8_F})=0.2776$,
and
$a({\bf 10})=0.4216$,
with
$\mu_u=1.8669$,
$\mu_d=-1.0256$,
and
$\mu_s=-0.6466$.
The predictions of this six parameter fit are displayed in the
\lq\lq Scalar sea" column of Table~\ref{tabla4}.

\paragraph*{\bf Vector sea.}
\label{iiib}

With seven parameters ($b_0$ and six $b(N)$'s) in the wavefunction and the
three $\mu_q$'s one can at best obtain a $\chi^2=34$ at the cost of
unrealistic values of $\mu_q$'s or $m_q$'s.
Unlike for the scalar sea, the pure vector sea modification of the wavefunction
is in the wrong direction.
This is probably because the vector sea contributions to $(\Delta q)^B$'s
has an overall opposite sign (see Table~\ref{tabla3}) to that of the scalar
sea contribution.

\paragraph*{\bf Scalar plus vector sea.}
\label{iiic}

All the parameters in the wavefunction of Eq.~(\ref{dos}) do not play a
significant role.
In fact, we find that excellent fits are obtained by describing the scalar sea
by only two parameters $a({\bf 8_F})$ and $a({\bf 10})$ and the vector sea by
one parameter either $b_0$ or $b({\bf 8_F})$.
Two fits to the magnetic moment data with six parameters only (3 for the sea
and 3 $\mu_q$'s) are listed in \lq\lq Case~1" and \lq\lq Case~2" columns of
Table~\ref{tabla4}.
For easy reference the parameters of the fits in Table~\ref{tabla4} are given
in Table~\ref{tabla5}.
Note that the values of the ratios $-\mu_u/2\mu_d\approx 0.95$ and
$\mu_s/\mu_d\approx 0.63$ for Cases~1 and 2 are practically the same as for SQM
fit of Ref.~\cite{10}.

\paragraph*{Case 1.}

The scalar sea is determined by
$a({\bf 8_F})=-0.1489$
and
$a({\bf 10})=0.4983$,
while the vector sea is described by the single parameter
$b({\bf 8_F})=0.5089$.
The values obtained for $\mu_q$'s are
$\mu_u=2.4550$,
$\mu_d=-1.2824$,
and
$\mu_s=-0.8071$,
implying
$m_u=254.79\ {\rm MeV}$,
$m_d=243.89\ {\rm MeV}$,
and
$m_s=387.51\ {\rm MeV}$.
The
$\chi^2/{\rm DOF}=1.43/2$.

\paragraph*{Case 2.}

The scalar sea parameters are again $a({\bf 8_F})$ and $a({\bf 10})$ with
values close to those of Case~1 while the vector sea is described by $b_0$
alone.
One finds
$a({\bf 8_F})=-0.1466$,
$a({\bf 10})=0.4932$,
and
$b_0=0.4779$,
with
$\mu_u=2.4748$,
$\mu_d=-1.3010$,
and
$\mu_s=-0.8249$.
The
$\chi^2/{\rm DOF}=2.23/2$.
The quark magnetic moments (masses) are slightly larger (smaller) than Case~1.

Comparison of the two fits reveals:
(a) In both cases, the two parameters $a({\bf 8_F})$ and $a({\bf 10})$
describing the scalar sea have practically the same values.
(b) Though, the vector sea is described by only one parameter its nature is
very different in the two cases.
In Case~1, the vector sea carries flavour (parameter $b({\bf 8_F})$) but in
Case~2 it is flavourless (parameter $b_0$).
(c) The $SU(3)$ breaking effects are solely due to the scalar sea parameter
$a({\bf 10})$.
Note that $a({\bf 10})$ contributes only to the wavefunction of the $\Sigma$'s
and $\Xi$'s.
Its inclusion dramatically improves the poor fit to $\mu_{\Sigma^{\pm}}$ and
$\mu_{\Xi^0}$ obtained for an octet physical baryon as mentioned above.
(d) In both cases the quark magnetic moments (masses) are larger (smaller) than
the SQM values but the ratios are in accord with SQM.
(e) It should be emphasized that the inclusion of more parameters to describe
the sea improves the fit only marginally.

In summary, the inclusion of a scalar sea or scalar plus vector sea in the
baryon give excellent fits to the spin 1/2 baryon magnetic moment data.
We feel our six parameters fits using actual experimental errors are
significant since most fits with four to five parameters invoke large
\lq\lq theoretical or notional errors" of a few percent to obtain reasonable
$\chi^2$-values.
It is gratifying that with just one more parameter we can fit the actual
experimental magnetic moment data and also are able to accomodate the EMC and
neutron beta decay data (see Secs.~\ref{iv} and \ref{vp}).

The magnetic moment fits determine all the $(\Delta q)^B$ which in turn have
implications for spin distrbutions and the $\Delta s=0$ axial vector weak decay
constant $g_A/g_V$.
We consider these for the nucleons and neutron beta decay where data are
available.

\section{Spin distributions}
\label{iv}

The spin distribution for the proton and neutron (in terms of quarks) are given
by

\begin{mathletters}
\label{trece}
\begin{equation}
I^p_1\equiv \int^1_0 g^p_1(x)dx =
\frac{1}{2}[\frac{4}{9}(\Delta u)^p + \frac{1}{9}(\Delta d)^p
+ \frac{1}{9}(\Delta s)^p],
\end{equation}
\begin{equation}
I^n_1\equiv \int^1_0 g^n_1(x)dx =
\frac{1}{2}[\frac{4}{9}(\Delta d)^p + \frac{1}{9}(\Delta u)^p
+ \frac{1}{9}(\Delta s)^p].
\end{equation}
\end{mathletters}

\noindent
The SQM values are
$I^p_1=5/18$
and
$I^n_1=0$
in disagreement with experiment~\cite{smc,burk} which gives
$I^p_1=0.126\pm 0.018$
and
$I^n_1=-0.08\pm 0.06$.
One must remark that the EMC experiment gives $I^p_1$ for
$\langle Q^2\rangle =10.7\ ({\rm Gev}/c)^2$ and this could be very different
for the very low $Q^2$ result predicted by our model or SQM.
Nevertheless, encouraged by the results of Sec.~\ref{results} we look at the
implication for $I^{p,n}_1$ using the magnetic moment fits for the
scalar sea and Cases~1 and 2.

\paragraph*{Scalar sea.}

For this fit
$(\Delta u)^p=1.3225$,
$(\Delta d)^p=-0.3045$,
and
$(\Delta s)^p=-0.0180$,
these give
$I^p_1=0.2760$
and
$I^n_1=0.0048$,
very close to the SQM values.
This is not surprising since for a pure scalar sea $\sum_q (\Delta q)^B=1$ for
each baryon independent of the number and values of the parameters describing
such a sea, that is, the baryon spin is carried entirely by the constituent
quarks.

The EMC experiment tells us that this is not so.
It is clear that for better fits to $I^{p,n}_1$ a vector sea is required so
that $\sum_q (\Delta q)^B\ne 1$, which is true for Cases~1 and 2.

\paragraph*{Case 1.}

Here
$(\Delta u)^p=0.9990$,
$(\Delta d)^p=-0.2602$,
and
$(\Delta s)^p=-0.0083$
giving
$\sum_q(\Delta q)^p=0.7305$.
These give
$I^p_1=0.2071$,
$I^n_1=-0.0028$,
with
$\chi^2=23$.

\paragraph*{Case 2.}

Here
$(\Delta u)^p=0.9999$,
$(\Delta d)^p=-0.2464$,
and
$(\Delta s)^p=0.0029$
giving
$\sum_q(\Delta q)^p=0.7564$.
These give
$I^p_1=0.2087$,
$I^n_1=0.0010$,
with
$\chi^2=25$.

It is clear that in both cases, quarks still carry a large fraction of the
proton spin and $\sum_q(\Delta q)^p$ is not small enough to give good agreement
for $I^p_1$.

Further combined fits to magnetic moments and EMC data with scalar and vector
sea were attempted.
A seven parameter fit yielded $\chi^2=1.83/3$, with
$a({\bf 8_F})=-0.26411$,
$a({\bf 10})=0.74254$,
$b({\bf 8_F})=0.70358$,
and
$b({\bf 8_D})=0.61006$
describing the sea, while
$\mu_u=3.4181$,
$\mu_d=-1.5076$,
and
$\mu_s=-0.9250$.
This fit yields
$(\Delta u)^p=0.6938$,
$(\Delta d)^p=-0.2616$,
and
$(\Delta s)^p=-0.0292$
so that
$\sum_q(\Delta q)^p=0.4030$
only.
This fit is not favoured as it gives a rather small value
($(\Delta u)^p-(\Delta d)^p=0.9554$) for $g_A/g_V$ for neutron decay discussed
below.

An experiment to measure the spin structure function of the $\Lambda$ has
been proposed recently~\cite{burk}.
Quark model gives

\begin{equation}
I^{\Lambda}_1\equiv \int^1_0 g^{\Lambda}_1(x)dx
=
\frac{1}{2}[
\frac{4}{9}(\Delta u)^{\Lambda}+
\frac{1}{9}(\Delta d)^{\Lambda}+
\frac{1}{9}(\Delta s)^{\Lambda}].
\label{catorce}
\end{equation}

\noindent
In SQM, this reduces to 1/18 ($=0.056$) since
$(\Delta u)^{\Lambda}=(\Delta d)^{\Lambda}=0$,
and
$(\Delta s)^{\Lambda}=1$.
For Case~1,
$(\Delta u)^{\Lambda}=(\Delta d)^{\Lambda}=-0.0083$,
and
$(\Delta s)^{\Lambda}=0.7472$
yielding the prediction
$I^{\Lambda}_1=0.0392$
which is lower than the SQM value.

\section{Axial vector weak decay constant ($\lowercase{g}_A/\lowercase{g}_V$)}
\label{vp}

Our general physical baryon wavefunction in Eq.~(\ref{dos}) respects isospin,
so that the physical $p$ and $n$ form an isodoublet and thus gives that for
neutron beta decay $n\to p+e^-+\bar{\nu}_e$

\begin{equation}
\frac{g_A}{g_V}=(\Delta u)^p-(\Delta d)^p.
\label{quince}
\end{equation}

\noindent
This is a very accurately measured quantity with the value $1.2573\pm 0.0028$.
Of the fits to the magnetic moments alone it is gratifying that the fits of
Case~1 and Case~2 automatically yield $g_A/g_V=1.2592$ and $g_A/g_V=1.2463$,
respectively.
This encouraged us to try a combined fit to the 11 pieces of data, viz.\ 8
magnetic moments, 2 spin distributions, and $g_A/g_V$.
The fits are very similar to those of Cases~1 and 2.
Of these, the best fit is practically the same as for Case~1 and differs from
it only in fitting $g_A/g_V$ more exactly and so is not displayed in
Table~\ref{tabla4}.
It gives a total $\chi^2/\mbox{DOF}=23.2/5$, where $I^{p,n}_1$ contribute
about 21.8 to the $\chi^2$ (which is also true for Case~1).

\section{Summary}
\label{v}

In summary, we have considered the physical spin 1/2 low-lying baryons to be
formed out of \lq\lq core" baryons (described by the $q^3$-wavefunction of SQM)
and a color singlet \lq\lq sea" which carries flavor and spin.
This sea (which may contain arbitrary number of gluons and $q\bar{q}$-pairs)
is specified only by its total flavor and spin quantum numbers.
The most general wavefunction for the physical baryons for an octet sea with
spin 0 and 1 was considered (Sec.~\ref{ii}) which respected isospin and
hypercharge (or strangeness).
Owing to the flavor properties of the sea the nucleons can have a non-zero
strange quark content (giving $(\Delta s)^p=(\Delta s)^n\neq 0$) through the
strange core baryons.
In this model the eight baryons no longer form an exact $SU(3)$ octet.
The admixture of other flavor $SU(3)$ representations in the wavefunction is
understood to represent broken $SU(3)$ effects.
The parameters in the wavefunction describing the sea were determined by
application to the baryon magnetic moment data, our primary objective.
We found an extremely good fit, with six parameters, to this data
{\em using available experimental errors}~\cite{10}.
Results are summarized in Tables~\ref{tabla4} and \ref{tabla5}.
Three of these parameters determined the sea contribution while the other three
were $\mu_q$'s (or $m_q$'s, $q=u, d, s$) the quark magnetic moments (masses).
The sea was found to be dominantly scalar (spin 0) described by 2
parameters while the admixture of the vector sea contributed only one
parameter.
The modified baryon wavefunction including such a sea which provides a good fit
to the magnetic moment data has only three parameters.
As a by product, for Case~1, we found a very good prediction for $g_A/g_V$ for
$n\to p+e^-+\bar{\nu}_e$.
The prediction for the spin distributions is in better agreement than SQM.
Our results (see Case~1 our best fit in Table~\ref{tabla4}) suggest that the
physical spin 1/2 \lq\lq octet" baryons contain an admixture of primarily the
{\bf 10} representation.
Why $SU(3)$ breaking (which we have invoked through a flavor octet sea) induce
these representations is a question for the future when one is able to
calculate
the parameters in the wavefunction of Eq.~(\ref{dos}) reliably from quantum
chromodynamics.

\acknowledgments

Special thanks are due to J.~Lach for his detailed and critical reading of
the manuscript.
This work was partially supported by CONACyT (M\'exico).

\mediumtext

\begin{table}
\caption{
Contribution to the physical baryon state $B(Y,I,I_3)$ formed out of
$\tilde{B}(Y,I,I_3)$ and flavor octet states $S(Y,I,I_3)$ (see third and
fourth terms in Eq.~(2)).
The core baryon states $\tilde{B}$ denoted by $\tilde{p}$, $\tilde{n}$, etc.
are the normal 3 valence quark states of SQM.
The sea octet states are denoted by $S_{\pi^+}=S(0,1,1)$, etc. as in Eq.~(4).
Further, $(\tilde{N}S_{\pi})_{I,I_3}$, $(\tilde{\Sigma}S_{\bar{K}})_{I,I_3}$,
$(\tilde{\Sigma}S_{\pi})_{I,I_3}$, $\dots$ stand for total $I$, $I_3$
{\em normalized} combinations of $\tilde{N}$ and $S_{\pi}$, etc.
See Table~II for the coefficients $\bar{\beta}_i$, $\beta_i$, $\gamma_i$,
and $\delta_i$.
}~
\label{tabla1}
\begin{tabular}
{
cc
}
$B(Y,I,I_3)$ &
$\tilde{B}(Y,I,I_3)$ and $S(Y,I,I_3)$
\\
\hline
\\
$p$ &
$
\bar{\beta}_1 \tilde{p} S_{\eta}
+ \bar{\beta}_2 \tilde{\Lambda} S_{K^+}
+ \bar{\beta}_3 (\tilde{N}S_{\pi})_{1/2,1/2}
+ \bar{\beta}_4 (\tilde{\Sigma}S_{K})_{1/2,1/2}
$
\\
\\
$n$ &
$
\bar{\beta}_1 \tilde{n} S_{\eta}
+ \bar{\beta}_2 \tilde{\Lambda} S_{K^0}
+ \bar{\beta}_3 (\tilde{N}S_{\pi})_{1/2,-1/2}
+ \bar{\beta}_4 (\tilde{\Sigma}S_{K})_{1/2,-1/2}
$
\\
\\
$\Xi^0$ &
$
\beta_1 \tilde{\Xi}^0 S_{\eta}
+ \beta_2 \tilde{\Lambda} S_{\bar{K}^0}
+ \beta_3 (\tilde{\Xi}S_{\pi})_{1/2,1/2}
+ \beta_4 (\tilde{\Sigma}S_{\bar{K}})_{1/2,1/2}
$
\\
\\
$\Xi^-$ &
$
\beta_1 \tilde{\Xi}^- S_{\eta}
+ \beta_2 \tilde{\Lambda} S_{\bar{K}^-}
+ \beta_3 (\tilde{\Xi}S_{\pi})_{1/2,-1/2}
+ \beta_4 (\tilde{\Sigma}S_{\bar{K}})_{1/2,-1/2}
$
\\
\\
$\Sigma^+$ &
$
\gamma_1 \tilde{p} S_{\bar{K}^0}
+ \gamma_2 \tilde{\Xi}^0 S_{K^+}
+ \gamma_3 \tilde{\Lambda} S_{\pi^+}
+ \gamma_4 \tilde{\Sigma}^+ S_{\eta}
+ \gamma_5 (\tilde{\Sigma}S_{\pi})_{1,1}
$
\\
\\
$\Sigma^-$ &
$
\gamma_1 \tilde{n} S_{K^-}
+ \gamma_2 \tilde{\Xi}^- S_{K^0}
+ \gamma_3 \tilde{\Lambda} S_{\pi^-}
+ \gamma_4 \tilde{\Sigma}^- S_{\eta}
+ \gamma_5 (\tilde{\Sigma}S_{\pi})_{1,-1}
$
\\
\\
$\Sigma^0$ &
$
\gamma_1 (\tilde{N} S_{\bar{K}})_{1,0}
+ \gamma_2 (\tilde{\Xi} S_{K})_{1,0}
+ \gamma_3 \tilde{\Lambda} S_{\pi^0}
+ \gamma_4 \tilde{\Sigma}^0 S_{\eta}
+ \gamma_5 (\tilde{\Sigma}S_{\pi})_{1,0}
$
\\
\\
$\Lambda$ &
$
\delta_1 (\tilde{N} S_{\bar{K}})_{0,0}
+ \delta_2 (\tilde{\Xi} S_{K})_{0,0}
+ \delta_3 \tilde{\Lambda} S_{\eta}
+ \delta_4 (\tilde{\Sigma}S_{\pi})_{0,0}
$
\\
\end{tabular}
\end{table}

\widetext

\begin{table}
\squeezetable
\caption{
The coefficients $\bar{\beta}_i$, $\beta_i$, $\gamma_i$, and $\delta_i$ in
Table~I expressed in terms of the coefficients $a(N)$,
$N={\bf 1, 8_{F}, 8_{D}, 10, \bar{10}, 27}$, in the $3^{\rm rd}$ term
(from scalar sea) in Eq.~(2).
The corresponding coefficients $\bar{\beta'}_i$, $\beta'_i$, $\gamma'_i$,
and $\delta'_i$ determining the flavor of structure of $4^{\rm th}$ term in
Eq.~(2) can be obtained from $\bar{\beta}_i$, etc. by the replacement
$a(N)\rightarrow b(N)$ (see text).
}~
\label{tabla2}
\begin{tabular}
{
cc
}
\\
$
\bar{\beta}_1=\frac{1}{\sqrt {20}}(3a({\bf 27})-a({\bf 8_D}))+
\frac{1}{2}(a({\bf 8_F})+a({\bf \bar{10}}))
$
&
$
\beta_1=\frac{1}{\sqrt {20}}(3a({\bf 27})-a({\bf 8_D}))-
\frac{1}{2}(a({\bf 8_F})-a({\bf 10}))
$
\\
\\
$
\bar{\beta}_2=\frac{1}{\sqrt {20}}(3a({\bf 27})-a({\bf 8_D}))-
\frac{1}{2}(a({\bf 8_F})+a({\bf \bar{10}}))
$
&
$
\beta_2=\frac{1}{\sqrt {20}}(3a({\bf 27})-a({\bf 8_D}))+
\frac{1}{2}(a({\bf 8_F})-a({\bf 10}))
$
\\
\\
$
\bar{\beta}_3=\frac{1}{\sqrt {20}}(a({\bf 27})+3a({\bf 8_D}))+
\frac{1}{2}(a({\bf 8_F})-a({\bf \bar{10}}))
$
&
$
\beta_3=-\frac{1}{\sqrt {20}}(a({\bf 27})+3a({\bf 8_D}))+
\frac{1}{2}(a({\bf 8_F})+a({\bf 10}))
$
\\
\\
$
\bar{\beta}_4 =-\frac{1}{\sqrt {20}}(a({\bf 27})+3a({\bf 8_D}))+
\frac{1}{2}(a({\bf 8_F})-a({\bf \bar{10}}))
$
&
$
\beta_4=\frac{1}{\sqrt{20}}(a({\bf 27})+3a({\bf 8_D}))+
\frac{1}{2}(a({\bf 8_F})+a({\bf 10}))
$
\\
\\
$
\gamma_1=\frac{1}{\sqrt{10}}(\sqrt{2}a({\bf 27})-\sqrt{3}a({\bf 8_D}))+
\frac{1}{\sqrt{6}}(a({\bf 8_F})-a({\bf 10})+a({\bf\bar{10}}))
$
&
$
\delta_1=\frac{1}{\sqrt{20}}(\sqrt{3}a({\bf 27})+\sqrt{2}a({\bf 8_D}))+
\frac{1}{2}(\sqrt{2}a({\bf 8_F})+a({\bf 1}))
$
\\
\\
$
\gamma_2=\frac{1}{\sqrt{10}}(\sqrt{2}a({\bf 27})-\sqrt{3}a({\bf 8_D}))-
\frac{1}{\sqrt{6}}(a({\bf 8_F})-a({\bf 10})+a({\bf\bar{10}}))
$
&
$
\delta_2=-\frac{1}{\sqrt{20}}(\sqrt{3}a({\bf 27})+\sqrt{2}a({\bf 8_D}))+
\frac{1}{2}(\sqrt{2}a({\bf 8_F})-a({\bf 1}))
$
\\
\\
$
\gamma_3=\frac{1}{\sqrt{10}}(\sqrt{3}a({\bf 27})+\sqrt{2}a({\bf 8_D}))-
\frac{1}{2}(a({\bf 10})+a({\bf\bar{10}}))
$
&
$
\delta_3=\frac{3\sqrt{3}}{\sqrt{40}}a({\bf 27})-
\frac{1}{\sqrt{5}}a({\bf 8_D})-
\frac{\sqrt{2}}{4}a({\bf 1})
$
\\
\\
$
\gamma_4=\frac{1}{\sqrt{10}}(\sqrt{3}a({\bf 27})+\sqrt{2}a({\bf 8_D}))+
\frac{1}{2}(a({\bf 10})+a({\bf\bar{10}}))
$
&
$
\delta_4=-\frac{1}{\sqrt{40}}a({\bf 27})-\sqrt{\frac{3}{5}}a({\bf 8_D})+
\frac{\sqrt{6}}{4}a({\bf 1})
$
\\
\\
$
\gamma_5=\frac{1}{\sqrt{6}}(2a({\bf 8_F})+a({\bf 10})-a({\bf\bar{10}}))
$
&
\\
\end{tabular}
\end{table}

\begin{table}
\squeezetable
\caption{
$(\Delta q)^B$ defined in Eq.~(8) for physical baryon $B$ given by general
wavefunction in Eq.~(2).
The normalizations $N_1$, $N_2$, $N_3$, and $N_4$ are given in Eqs.~(6).
The $(\Delta q)^{\Sigma^0\Lambda}$ for the $\Sigma^0\rightarrow\Lambda$
transition magnetic moment is also given.
}~
\label{tabla3}
\begin{tabular}
{
c
}
\\
$
(\Delta u)^p=
\frac{1}{3N^2_1}
[
4(1-\frac{1}{3}b^2_0)
+
(
4\bar{\beta}^2_1+\frac{2}{3}\bar{\beta}^2_3+\frac{10}{3}\bar{\beta}^2_4
-2\bar{\beta}_2\bar{\beta}_4
)
-\frac{1}{3}
(
\bar{\beta}_i\rightarrow\bar{\beta'}_i
)
]
$
\\
\\
$
(\Delta d)^p=
\frac{1}{3N^2_1}
[
-(1-\frac{1}{3}b^2_0)
+
(
-\bar{\beta}^2_1+\frac{7}{3}\bar{\beta}^2_3+\frac{2}{3}\bar{\beta}^2_4
+2\bar{\beta}_2\bar{\beta}_4
)
-\frac{1}{3}
(
\bar{\beta}_i\rightarrow\bar{\beta'}_i
)
]
\ \ \ \ \ \
(\Delta s)^p=
\frac{1}{3N^2_1}
[
(
3\bar{\beta}^2_2-\bar{\beta}^2_4
)
-\frac{1}{3}
(
\bar{\beta}_i\rightarrow\bar{\beta'}_i
)
]
$
\\
\\
$
(\Delta u)^n=(\Delta d)^p
\ \ \ \ \ \
(\Delta d)^n=(\Delta u)^p
\ \ \ \ \ \
(\Delta s)^n=(\Delta s)^p
$
\\
\\
$
(\Delta u)^{\Xi^0}=
\frac{1}{3N^2_2}
[
-(1-\frac{1}{3}b^2_0)
+
(
-\beta^2_1-\frac{1}{3}\beta^2_3+\frac{10}{3}\beta^2_4
-2\beta_2\beta_4
)
-\frac{1}{3}
(
\beta_i\rightarrow\beta'_i
)
]
$
\\
\\
$
(\Delta d)^{\Xi^0}=
\frac{1}{3N^2_2}
[
(
-\frac{2}{3}\beta^2_3+\frac{2}{3}\beta^2_4
+2\beta_2\beta_4
)
-\frac{1}{3}
(
\beta_i\rightarrow\beta'_i
)
]
\ \ \ \ \ \
(\Delta s)^{\Xi^0}=
\frac{1}{3N^2_2}
[
4(1-\frac{1}{3}b^2_0)
+
(
4\beta^2_1+3\beta^2_2+4\beta^2_3-\beta^2_4
)
-\frac{1}{3}
(
\beta_i\rightarrow\beta'_i
)
]
$
\\
\\
$
(\Delta u)^{\Xi^-}=(\Delta d)^{\Xi^0}
\ \ \ \ \ \
(\Delta d)^{\Xi^-}=(\Delta u)^{\Xi^0}
\ \ \ \ \ \
(\Delta s)^{\Xi^-}=(\Delta s)^{\Xi^0}
$
\\
\\
$
(\Delta u)^{\Sigma^+}=
\frac{1}{3N^2_3}
[
4(1-\frac{1}{3}b^2_0)
+
(
4\gamma^2_1-\gamma^2_2+4\gamma^2_4+3\gamma^2_5
-\sqrt{6}\gamma_3\gamma_5
)
-\frac{1}{3}
(
\gamma_i\rightarrow\gamma'_i
)
]
$
\\
\\
$
(\Delta d)^{\Sigma^+}=
\frac{1}{3N^2_3}
[
(
-\gamma^2_1+\gamma^2_5
+\sqrt{6}\gamma_3\gamma_5
)
-\frac{1}{3}
(
\gamma_i\rightarrow\gamma'_i
)
]
\ \ \ \ \ \
(\Delta s)^{\Sigma^+}=
\frac{1}{3N^2_3}
[
-(1-\frac{1}{3}b^2_0)
+
(
4\gamma^2_2+3\gamma^2_3-\gamma^2_4-\gamma^2_5
)
-\frac{1}{3}
(
\gamma_i\rightarrow\gamma'_i
)
]
$
\\
\\
$
(\Delta u)^{\Sigma^-}=(\Delta d)^{\Sigma^+}
\ \ \ \ \ \
(\Delta d)^{\Sigma^-}=(\Delta u)^{\Sigma^+}
\ \ \ \ \ \
(\Delta s)^{\Sigma^-}=(\Delta s)^{\Sigma^+}
$
\\
\\
$
(\Delta u)^{\Sigma^0}=
\frac{1}{2}[(\Delta u)^{\Sigma^+}+(\Delta u)^{\Sigma^-}]
\ \ \ \ \ \
(\Delta d)^{\Sigma^0}=(\Delta u)^{\Sigma^0}
\ \ \ \ \ \
(\Delta s)^{\Sigma^0}=(\Delta s)^{\Sigma^+}
$
\\
\\
$
(\Delta u)^{\Lambda}=
\frac{1}{3N^2_4}
[
(
\frac{3}{2}\delta^2_1-\frac{1}{2}\delta^2_2+2\delta^2_4
)
-\frac{1}{3}
(
\delta_i\rightarrow\delta'_i
)
]
$
\\
\\
$
(\Delta d)^{\Lambda}=(\Delta u)^{\Lambda}
\ \ \ \ \ \ \ \ \
(\Delta s)^{\Lambda}=
\frac{1}{3N^2_4}
[
3(1-\frac{1}{3}b^2_0)
+
(
4\delta^2_2+3\delta^2_3-\delta^2_4
)
-\frac{1}{3}
(
\delta_i\rightarrow\delta'_i
)
]
$
\\
\\
$
(\Delta u)^{\Sigma^0\Lambda}=
\frac{1}{N_3N_4}
[
\frac{1}{\sqrt{3}}(1-\frac{1}{3}b^2_0)
+
(
\frac{1}{\sqrt{3}}\gamma_4\delta_3-\frac{1}{3}\gamma_3\delta_4
+
\frac{5}{6}\gamma_1\delta_1-\frac{1}{6}\gamma_2\delta_2
+\frac{4}{3\sqrt{6}}\gamma_5\delta_4
)
-\frac{1}{3}
(
\gamma_i,\delta_i\rightarrow\gamma'_i,\delta'_i
)
]
$
\\
\\
$
(\Delta d)^{\Sigma^0\Lambda}=-(\Delta u)^{\Sigma^0\Lambda}
\ \ \ \ \ \ \ \ \ \ \ \
(\Delta s)^{\Sigma^0\Lambda}=0
$
\\
\end{tabular}
\end{table}

\begin{table}
\squeezetable
\caption{
Results for the baryon magnetic moments (in N.M.), spin distributions,
and $g_A/g_V$ along with available data and SQM fit for comparison.
In each case, the $\chi^2/\mbox{DOF}$ quoted is for the fit to magnetic moment
data alone.
Last three columns give our fits using six parameters: 3 $\mu_q$'s and 3
parameters for the sea.
Column~4 gives the results for a pure scalar sea while the last two columns
give those for scalar sea (2 parameters) plus vector sea (1 parameter) labelled
as Cases~1 and 2.
The predictions for $I^{p,n}_1$ and $g_A/g_V$ (from the fit to magnetic
moments) is given in the last three rows.
Case~1 gives the best fit.
For details see text and Table~V where the fit parameters are compared.
}~
\label{tabla4}
\begin{tabular}
{
c
r@{\,$\pm$\,}l
d
d
d
d
}
Baryon &
\multicolumn{2}{c}{Data} &
SQM &
Scalar sea &
Case 1 &
Case 2
\\
\hline
\\
$p$ &
2.79284739 & $6\times 10^{-8}$ &
2.7928 &
2.7928 &
2.7928 &
2.7928
\\
\\
$n$ &
$-$1.9130428 & $5\times 10^{-7}$ &
$-$1.9130 &
$-$1.9130 &
$-$1.9130 &
$-$1.9130
\\
\\
$\Lambda$ &
$-$0.613 & 0.004 &
$-$0.701 &
$-$0.616 &
$-$0.613 &
$-$0.616
\\
\\
$\Sigma^+$ &
2.458 & 0.010 &
2.703 &
2.456 &
2.457 &
2.459
\\
\\
$\Sigma^0$ &
\multicolumn{2}{c}{--------} &
0.8203 &
0.6238 &
0.6371 &
0.6362
\\
\\
$\Sigma^-$ &
$-$1.160 & 0.025 &
$-$1.062 &
$-$1.208 &
$-$1.183 &
$-$1.186
\\
\\
$\Xi^0$ &
$-$1.250 & 0.014 &
$-$1.552 &
$-$1.248 &
$-$1.249 &
$-$1.253
\\
\\
$\Xi^-$ &
$-$0.6507 & 0.0025 &
$-$0.6111 &
$-$0.6500 &
$-$0.6504 &
$-$0.6507
\\
\\
$|\Sigma^0\rightarrow\Lambda|$ &
1.61 & 0.08 &
1.63 &
1.52 &
1.55 &
1.55
\\
\\
$\chi^2/\mbox{DOF}$ &
\multicolumn{2}{c}{--------} &
1818/5 &
5.60/2 &
1.43/2 &
2.23/2
\\
\\
$I^p_1$ &
0.126 & 0.018 &
0.278 &
0.276 &
0.207 &
0.209
\\
\\
$I^n_1$ &
$-$0.08 & 0.06 &
0 &
0.005 &
$-$0.003 &
0.001
\\
\\
$g_A/g_V$ &
1.2573 & 0.0028 &
1.6667 &
1.6270 &
1.2592 &
1.2463
\\
\end{tabular}
\end{table}

\mediumtext

\begin{table}
\squeezetable
\caption{
Comparison of the 6 parameters for various fits to the magnetic moments and the
values of $(\Delta q)^p$ obtained.
The SQM values for $\mu_q$ and $(\Delta q)^p$ are also given.
The parameters $a({\bf 8_D})$, $a({\bf 8_F})$, and $a({\bf 10})$ refer to a
scalar sea while $b_0$ and $b({\bf 8_F})$ refer to a vector sea.
See text for details and Table~IV for results.
}~
\label{tabla5}
\begin{tabular}
{
c
d
d
d
d
}
&
SQM &
Scalar sea &
\multicolumn{2}{c}{Scalar plus vector sea}
\\
&
&
&
Case 1 &
Case 2
\\
\hline
\\
$\mu_u$ &
1.8517 &
1.8669 &
2.4550 &
2.4748
\\
\\
$\mu_d$ &
$-$0.9719 &
$-$1.0256 &
$-$1.2824 &
$-$1.3010
\\
\\
$\mu_s$ &
$-$0.7013 &
$-$0.6466 &
$-$0.8071 &
$-$0.8249
\\
\\
$a({\bf 8_D})$ &
\multicolumn{1}{r}{------} &
$-$0.2262 &
\multicolumn{1}{r}{------} &
\multicolumn{1}{r}{------}
\\
\\
$a({\bf 8_F})$ &
\multicolumn{1}{r}{------} &
0.2776 &
$-$0.1489 &
$-$0.1466
\\
\\
$a({\bf 10})$ &
\multicolumn{1}{r}{------} &
0.4216 &
0.4983 &
0.4932
\\
\\
$b({\bf 8_F})$ &
\multicolumn{1}{r}{------} &
\multicolumn{1}{r}{------} &
0.5089 &
\multicolumn{1}{r}{------}
\\
\\
$b_0$ &
\multicolumn{1}{r}{------} &
\multicolumn{1}{r}{------} &
\multicolumn{1}{r}{------} &
0.4779
\\
\\
$(\Delta u)^p$ &
4/3 &
1.3225 &
0.9990 &
0.9999
\\
\\
$(\Delta d)^p$ &
$-$1/3 &
$-$0.3045 &
$-$0.2602 &
$-$0.2464
\\
\\
$(\Delta s)^p$ &
0 &
$-$0.0180 &
$-$0.0083 &
0.0029
\\
\end{tabular}
\end{table}

\end{document}